\documentclass[a4paper]{jpconf}

\usepackage{amssymb,amsmath,graphicx,color,caption,subcaption,soul}
\usepackage[colorlinks=true,urlcolor=black]{hyperref}

\hypersetup{backref = true,pagebackref = true,hyperindex = true, colorlinks = true,
breaklinks = true, urlcolor = blue, linkcolor = blue, bookmarks = true,
bookmarksopen = true, citecolor=red}

\newcommand{\R}{\mathbb{R}}

\def\eps{\varepsilon}

\def\D{{\rm d}}
\def\be{\begin{equation}}
\def\ee{\end{equation}}

\begin{document}

\title{Flat polymerized membranes at three-loop order}

\author{S.~Metayer$^{1\,*}$, D.~Mouhanna$^{2\,\dag}$, S.~Teber$^{1\,\ddag}$}

\date{December 2021}

\address{$^1$ Sorbonne Universit\'e, CNRS, Laboratoire de Physique Th\'eorique et Hautes Energies, LPTHE, F-75005 Paris, France}
\address{$^2$ Sorbonne Universit\'e, CNRS, Laboratoire de Physique Théorique de la Matière Condensée, LPTMC, F-75005 Paris, France}

\ead{$^*$smetayer@lpthe.jussieu.fr, $^\dag$mouhanna@lptmc.jussieu.fr, $^\ddag$teber@lpthe.jussieu.fr}

\begin{abstract}
In this conference report, we present a recent field theoretic renormalization group analysis of flat polymerized membranes at three-loop order by the present authors [Phys. Rev. E {\bf 105}, L012603 (2022)].
\end{abstract}

\section{Context}

In this conference report, we present a brief overview of the methods and the main results of Ref.~\cite{Metayer:2021}. The latter focuses on a field theoretic study of the flat phase of polymerized membranes, see, e.g., \cite{Nelson:1987,Aronovitz:1988,Guitter:1988,Aronovitz:1989,Guitter:1989,Guitter:1990} 
for early studies. Such a model has been the subject of extensive studies during the last decades especially that it seems to adequately describe the elastic degrees of freedom of graphene and graphene-like materials, see, e.g., the textbook \cite{katsnelson12}. A major challenge in this context is an accurate determination of the renormalization group (RG) functions of the model and in particular the field anomalous dimension $\eta$. This critical exponent, alone, determines all the power-law scaling behaviors of the theory in the infra-red (IR) regime leading to an anomalous rigidity characteristic of the flat phase, see, e.g., \cite{LeDoussal:2018} for a review.

Early, one-loop, computations were carried out in the seminal work of Aronovitz and Lubensky \cite{Aronovitz:1988} and led to $\eta=0.96$. Due to major computational challenges, over 30 years passed before the achievement of the full two-loop computations in \cite{Coquand:2020} that made use of massless multi-loop techniques (see, e.g., \cite{Kotikov:2019} for a review) and led to $\eta=0.9139$. During this time, a lot of other approaches have been carried out, including non-perturbative techniques such as NPRG \cite{Kownacki:2009} and SCSA \cite{LeDoussal:1992,LeDoussal:2018} as well as numerical Monte-Carlo techniques \cite{Zhang:1993,Bowick:1996,Troster:2013,Los:2009}. All these methods rely on their own sets of approximations and have led to scattered values for $\eta$ in the range $[0.72,0.88]$.
Very recently, the field theoretic approach could be extended to three-loop order in \cite{Metayer:2021}
thanks to a full state of the art automation of the computations using packages originally developed for  high-energy physics calculations \cite{Nogueira:1993,Lee:2014}. 

In the following, we will first introduce the model of a flat polymerized membrane. Then, we will briefly review the field theoretic RG 
 approach used in our calculations. Finally, we will present our results and conclude. 

\section{Model}

    We consider a $D$-dimensional homogeneous and isotropic membrane embedded in a $d$-dimensional space. Each mass point of the membrane is indexed by $\vec x \in \R^D$. In $\R^d$, the reference state of the membrane is the unperturbed flat state where each of these mass points is indexed by $\vec R^{(0)}(\vec x\,) = (\vec x, \vec 0_{d_c})$ where $\vec 0_{d_c}$ is the null vector of co-dimension $d_c = d-D\ge0$. Allowing for small displacements inside the membrane, the latter are parameterized by a phonon field $\vec u(\vec x\,) \in \R^D$ and a flexuron field $\vec h(\vec x\,) \in \R^{d_c}$ such that the perturbed mass points are located at: $\vec R(\vec x\,) = (\vec x + \vec u(\vec x\,), \vec h(\vec x\,))$. The induced metric is then defined as: 
    $g_{ij} = \partial_i \vec R(\vec x\,) \cdot \partial_j \vec R(\vec x\,)$ ($g_{ij}^{(0)}=\delta_{ij}$ in the unperturbed state). The strain tensor
    is defined as: 
    \be
        u_{ij} = \frac{1}{2}\, \left( \partial_i \vec R(\vec x\,) \cdot \partial_j \vec R(\vec x\,) - g_{ij}^{(0)} \right) = \frac{1}{2}\,\left(\partial_i u_j + \partial_j u_i + \partial_i \vec h \cdot \partial_j \vec h + \partial_i \vec u \cdot \partial_j \vec u \right) \, .
    \ee

    The Euclidean low-energy action of the membrane reads \cite{Nelson:1987,Aronovitz:1988,Guitter:1988,Aronovitz:1989,Guitter:1989,Guitter:1990}:
    \be
        S[\vec u, \vec h\,] = \int\!\D^D x \left[ \frac{1}{2}\,(\Delta \vec h)^2 + \frac{\lambda}{2}\,u_{aa}^2 + \mu\,u_{ab}^2 \right]\, ,
        \label{eq:S}
    \ee
    where $\lambda$ and $\mu$ are Lam\'e elastic moduli. Moreover, the bending rigidity has been set to $\kappa=1$ without restricting the generality of the problem.
    
    Upon neglecting nonlinearities in the phonon field $\vec{u}$, the action (\ref{eq:S}) yields the so-called {\it two-field model} that involves cubic and quartic interactions in the fields $\vec{u}$ and $\vec{h}$ and coupling constants $\lambda$ and $\mu$. Because this model is quadratic in the phonon field, the latter can be integrated over exactly. This then yields another, equivalent, model, the so-called {\it flexural effective model}, that depends only on the flexural field $\vec h$ and on coupling constants $\lambda$ and $b$, see \cite{Metayer:2021} for more details.
   
    Both of these models are highly derivative field theories that, after Fourier transform, result in non-trivial momentum dependencies of the Feynman rules, and ultimately, to very lengthy expressions for the Feynman diagrams. They also display rather non-trivial tensorial structures that need to be projected out carefully especially for the effective model. The massless nature of both models translates the long-range nature of elastic interactions that is responsible for the renormalization of the two sets  of couplings $(\lambda,\mu)$ and $(b,\mu)$. Fortunately, it allows to apply powerful multiloop techniques,  see  the review \cite{Kotikov:2019}. Moreover, this theory is highly constrained by Ward identities that relate all renormalization constants to two-point correlation functions only, which further reduces the complexity of the problem.

\section{Perturbative renormalization group approach}
    
    In \cite{Metayer:2021}, we have analysed the RG 
     flows of both the two-field and the flexural effective models up to three-loop order. Following the two-loop computations of  \cite{Coquand:2020}, the three-loop case has been achieved in \cite{Metayer:2021} thanks to a full automation of the calculations. The latter can be summarized in the following steps:
    \begin{enumerate}
        \item Compute the bare flexuron self-energy $\Sigma$ and the bare polarization $\Pi$ from Feynman diagram expansions.\footnote{\label{footnote:Pi} The polarization $\Pi$ corresponds to the phonon polarization in the two-field model (with longitudinal $\Pi_\parallel$ and transverse $\Pi_\bot$ projections) and to an effective polarization in the effective model (with projections $\Pi_N$ and $\Pi_M$ on irreducible tensors $N$ and $M$ that are defined, e.g., in the review \cite{LeDoussal:2018}).}  These computations were carried out for a membrane with arbitrary codimension $d_c$ using 
         dimensional regularization in $D=4-2\eps$.\footnote{Note that the orginal paper \cite{Metayer:2021} uses the convention $D=4-\eps$.} A total of $61$ distinct diagrams had to be computed in the two-field model and $32$ distinct diagrams in the effective flexural model, see Table \ref{tab:numberofdiags}. The computational time was approximately the same for the two sets of diagrams due to the intricate tensorial  structure of the effective model. Diagrams were generated using \textsc{Qgraf} \cite{Nogueira:1993} and then imported to \textsc{Mathematica} where the numerator algebra and tensor manipulations were performed with home made codes. The reduction to master integrals was automated with the program \textsc{LiteRed} \cite{Lee:2014} and the analytic expression of some complicated masters taken from, e.g., the review \cite{Kotikov:2019}.
        \begin{table}[h]
            \caption{Number of distinct diagrams in the two-field and effective flexural models}
            \label{tab:numberofdiags}
            \centering
            \begin{subtable}[h]{0.43\textwidth}
                \centering
                \begin{tabular}{cccc}
                    \br
                    2-field & 1-loop & 2-loop & 3-loop \\
                    \mr
                    $\Sigma$ & 1 & 5 & 32 \\
                    $\Pi$ & 1 & 3 & 19 \\
                    \br
                \end{tabular}
            \end{subtable}
            \begin{subtable}[h]{0.43\textwidth}
                \centering
                \begin{tabular}{cccc}
                    \br
                    effective & 1-loop & 2-loop & 3-loop \\
                    \mr
                    $\Sigma$ & 1 & 3 & 15 \\
                    $\Pi$ & 1 & 2 & 11 \\
                    \br
                \end{tabular}
            \end{subtable}
        \end{table}
        \item Compute the renormalization constants $Z$, $Z_\lambda$, $Z_\mu$ in the two-field model and $Z$, $Z_b$, $Z_\mu$ in the effective model. They are defined as: $h=Z^{1/2}\,h_r$ (together with $u=Z\,u_r$ in the two-field model), $\lambda=M^{2\eps}Z_\lambda\,\lambda_r$,  $b=M^{2\eps}Z_b\,b_r$ and $\mu=M^{2\eps}Z_\mu\,\mu_r$ where $M$ is the renormalization mass scale. The constants were found with the help of simple algebraic relations (with no need to compute any additional counter-term diagrams) that read
        \be
            \text{finite}=(p^4-\Sigma)\, Z,\quad
            \text{finite}=(p^2\,Z_\mu\,\mu_r-\Pi_\bot)\,Z^2,\quad
            \text{finite}=(p^2(Z_\lambda\,\lambda_r + 2\,Z_\mu\,\mu_r)-\Pi_\parallel)\,Z^{2}\, ,
        \ee
        for the two-field model and
        \be
            \text{finite}=(p^4-\Sigma)\,Z,\quad
            \text{finite}=((Z_\mu\,\mu_r)^{-1}-\Pi_{M})\,Z^{-2},\quad
            \text{finite}=((Z_b\,b_r)^{-1}-\Pi_{N})\,Z^{-2}\, ,
        \ee
        for the effective model, where $\text{finite}$ means of order $\eps^0$.
        
        %
        
        \item Compute the RG functions in both models, consisting in the beta functions $\beta_x = M \, \partial_M Z_x$ ($x=\{\lambda,\mu,b\}$) and the field anomalous dimension $\eta = M \, \partial_M \log Z$. Note that, at this point, these functions differ in the two models.
        \item Solve perturbatively the system of beta functions: $\{\,\beta_\lambda(\lambda^*,\mu^*)=0\,,\,\,\beta_\mu(\lambda^*,\mu^*)=0 \,\}$ in the two-field model and $\{\,\beta_\mu(\mu^*,b^*)=0\,,\,\,\beta_b(\mu^*,b^*)=0\,\}$ in the effective flexural model. This allows
        to access the different scale invariant fixed points: ($\lambda^*,\mu^*$) in the two-field model and ($\mu^*,b^*$) in the effective model. Once again, the expressions of the fixed points differ in the two models.
        \item Compute the anomalous dimension $\eta$ at the various 
         fixed-points for each model. A strong check of our calculations is that, for the special case of the non-trivial and IR stable fixed point (see more below), both $\eta(\lambda^*,\mu^*)$ in the two-field model and $\eta(\mu^*,b^*)$ in the effective model are equal. This is in accordance with the fact that the two models are identical and that these quantities are scheme-independent and universal.    
    \end{enumerate}

\section{Results}
    
    We now proceed on summarizing the results of the above calculations. In both models, four perturbative fixed points are obtained. In the case of the two-field model, they correspond to: 
    \begin{enumerate}
        \item P$_1$: the unstable gaussian  fixed-point ($\lambda_1^*=\mu_1^*=0$) with $\eta_1=0$,
        
        \item P$_2$: the unstable shearless  fixed-point ($\lambda_2^*=32\pi^2\eps/d_c$, $\mu_2^*=0$) with $\eta_2=0$,
        
        \item P$_3$: this fixed point has non-trivial values of $\lambda_3^*$ and $\mu_3^*$ (see \cite{Metayer:2021}) that lead to a strictly negative bulk-modulus at three-loop ($B_3^*=\lambda_3^*+2\mu_3^*/D < 0$); it therefore lies outside of the (mechanical) stability region of the model ($B \geq 0$). For the sake of comparison with the effective model, we nevertheless provide the expression of the corresponding anomalous dimension in the case $d_c=1$:
        \be
        \eta_3=0.9524\, \eps -0.0711\, \eps^2-0.0698\, \eps^3 +O(\eps^4) \, ,
        \label{etaP3}
        \ee
        \item P$_4$: the IR-stable non-trivial fixed point (see \cite{Metayer:2021} for the expressions of $\lambda_4^*$ and $\mu_4^*$) with $\eta_4$ provided below in the physical case $d_c=1$, see (\ref{eq:etaseries}).
    \end{enumerate}
    In the case of the effective flexural model, the fixed points correspond to:
    \begin{enumerate}
        \item P$_1$: the unstable gaussian fixed-point ($\mu_1^*=b_1^*=0$)  with $\eta_1=0$,
        
        \item P$_2'$: an unstable shearless fixed-point ($\mu_2^*=0$) with a non-trivial expression for $b_2^*$ \cite{Metayer:2021}. Contrary to P$_2$, this fixed point has a non trivial anomalous dimension that reads (in the case $d_c=1$):
        \begin{equation}
        \eta_2'= 0.8000\, \eps -0.0053 \, \eps^2 + 0.0110\, \eps^3 + O(\eps^4)\, ,
        \label{seta2'pert}
        \end{equation} 
        \item P$_3$: the infinitely compressible fixed point with $b_3^*=0$ and a non-trivial expression for $\mu_3^*$ \cite{Metayer:2021}. Contrary to the case of the two-field model, it now has a vanishing bulk modulus ($B_3^*=0$) and is therefore located on the (mechanical) stability line of the model. Interestingly, the corresponding $\eta_3$ in the effective model corresponds exactly to $\eta_3$ in the two-field model, see (\ref{etaP3}) in the case $d_c=1$.
        
        \item P$_4$: the IR-stable non-trivial fixed point (see \cite{Metayer:2021} for the expressions of $\mu_4^*$ and $b_4^*$) with $\eta_4$ exactly equal to the one found in the two-field model and provided below in the physical case $d_c=1$, see (\ref{eq:etaseries}).
    \end{enumerate}
    
    From the above results, we see that the globally IR attractive fixed point that controls the physics of the flat phase is P$_4$ at three-loops (as was already the case at one- and two-loops). Both the two-field model and the effective flexural model yield the same field anomalous dimension at this  fixed point. Referring to it simply as $\eta$, its expression in the physical case $d_c=1$ reads:
    \be
        \eta \equiv \eta_4=\frac{24}{25}\,\eps
        -\frac{144}{3125}\,\eps^2
        -\frac{4(1286928\,\zeta_3-568241)}{146484375}\,\eps^3+O(\eps^4)\, ,
        \label{eq:etaseries}
    \ee
    with $\zeta_3\approx1.202$ being the Ap\'ery constant. Numerically evaluating the coefficients yields:
    \be
    \eta = 0.9600\, \eps -0.0461\, \eps^2-0.0267\, \eps^3 + O(\eps^4)\, .
\label{seta4pert}
    \ee
    Interestingly, the coefficients of (\ref{seta4pert}) are small and even decreasing with increasing loop order.\footnote{This is due to the structure of the series that consists of very large denominators, see \cite{Metayer:2021} for details.} The asymptotic series seems to be in a convergent regime for which an extrapolation to the case of interest ($\eps=1$) does not require any resummation. Order by order in the perturbative expansion, we therefore have (for $d_c=1$ and $D=2$):
    \be
        \eta_{\text{1-loop}}=0.96~\cite{Aronovitz:1988}\,,\quad
        \eta_{\text{2-loop}}\approx 0.9139~\cite{Coquand:2020}\,,\quad
        \eta_{\text{3-loop}}\approx 0.8872~\cite{Metayer:2021}\,.
    \ee
    Clearly, with increasing loop order, the result gets closer to the range of values obtained by other methods discussed in the introduction; $[0.72,0.88]$.
    \begin{table}[t]
        \caption{Benchmarking other approaches using $\eta$}
        \label{tab:benchmark}
        \centering
        \begin{tabular}{ccc}
            \br
            $\eta$ & $\eps=1$ & (Re)-expanded in $\eps$ \\
            \mr
            3-loop \cite{Metayer:2021} & $0.8872$ & $0.96\,\eps-0.0461\,\eps^2-0.0267\,\eps^3$ \\
            SCSA \cite{LeDoussal:1992,LeDoussal:2018} & $0.8209$ & $0.96\,\eps-0.0476\,\eps^2-0.0280\,\eps^3$ \\
            NPRG \cite{Kownacki:2009} & $0.8491$ & $0.96\,\eps-0.0367\,\eps^2-0.0266\,\eps^3$ \\
            \br
        \end{tabular}
    \end{table}

    Because our results for $\eta$ are exact order by order and universal, we may use them as a benchmark for other methods, such as NPRG and SCSA. This is provided by Table \ref{tab:benchmark} that displays a striking proximity between all approaches. Note that more detailed comparisons involving all of the fixed points can be found in \cite{Metayer:2021}.
    
\section{Discussion}

    In this short conference report, we have reviewed recent calculations of RG functions in two equivalent models of flat polymerized membranes at three-loop order \cite{Metayer:2021}. A full state of the art automation of all tasks was necessary to achieve such a goal as the corresponding field theories are highly derivative with rather intricate tensor structures. 
    
    A central result was the computation of
    the field anomalous dimension $\eta$ at the globally attractive fixed point controlling the physics of the flat phase in $D=4-2\eps$ and for arbitrary $d_c$. A strong check of our results was that the expression for $\eta$ was found to be the same in both models (despite the fact that intermediate steps differ). The associated (asymptotic) $\eps$-series displayed remarkably small and decreasing coefficients with increasing loop order, see (\ref{seta4pert}). Without any resummation needed, we found $\eta=0.8872$ for the physical case $D=2$ and $d_c=1$. This result lies in the range of values obtained by other methods $[0.72,0.88]$. It also revealed the striking ability of both SCSA and NPRG to numerically mimic the true perturbative expansion (Table \ref{tab:benchmark}).
    
    Note: while writing this conference proceedings, the four-loop computation, in the two-field model, was achieved by Pikelner \cite{Pikelner:2021} using similar techniques, thereby confirming our three-loop results and leading to the new improved value $\eta_{\text{4-loop}}=0.8670$. 

\ack

    S.~Metayer thanks the organizers of the conference ``Advanced Computing and Analysis Techniques'' (ACAT) 2021 for providing the opportunity to present this work through two poster sessions, which lead to fruitful discussions with participants.

\section*{References}

    \bibliography{main}
    \bibliographystyle{iopart-num}

\end{document}